\documentstyle[12pt]{article}
\begin{document}
\renewcommand{\thesection}{\Roman{section}}
\renewcommand{\thetable}{\Roman{table}}
\newcommand{\be}{\begin{eqnarray}}
\newcommand{\beq}{\begin{equation}}
\newcommand{\ba}{\begin{array}}
\newcommand{\ee}{\end{eqnarray}}
\newcommand{\eeq}{\end{equation}}
\newcommand{\ea}{\end{array}}
\newcommand{\zt}{\zeta}
\newcommand{\ve}{\varepsilon}
\newcommand{\al}{\alpha}
\newcommand{\gm}{\gamma}
\newcommand{\Gm}{\Gamma}
\newcommand{\om}{\omega}
\newcommand{\et}{\eta}
\newcommand{\bt}{\beta}
\newcommand{\dt}{\delta}
\newcommand{\Dt}{\Delta}
\newcommand{\La}{\Lambda}
\newcommand{\la}{\lambda}
\newcommand{\vp}{\varphi}
\newcommand{\nn}{\nonumber}
\newcommand{\nid}{\noindent}

\begin{titlepage}
\begin{center}
 {\Large \bf Critical behavior of certain antiferromagnets with complicated
 ordering: Four-loop $\ve$-expansion analysis}
\vspace{1cm}

 {\large A. I. Mudrov\footnote{E-mail address: mudrova@macs.biu.ac.il}} \\
 {\it Department of Mathematics, Bar-Ilan University,
      52900 Ramat-Gan, Israel} \\

\bigskip
 {\large K. B. Varnashev\footnote{E-mail address: kvarnash@kv8100.spb.edu}} \\
 {\it Department of Physical Electronics, Saint Petersburg
      Electrotechnical University, \\
      Professor Popov Street  5, St. Petersburg, 197376, Russia}
\end{center}

\begin{abstract}
\vspace{0.25cm}

\nid
The critical behavior of a complex $N$-component order parameter
Ginzburg-Landau model with isotropic and cubic interactions describing
antiferromagnetic and structural phase transitions in certain crystals
with complicated ordering is studied in the framework of the four-loop
renormalization group (RG) approach in $(4-\ve)$ dimensions. By using
dimensional regularization and the minimal subtraction scheme, the
perturbative expansions for RG functions are deduced and resummed
by the Borel-Leroy transformation combined with a conformal mapping.
Investigation of the global structure of RG flows for the physically
significant cases $N=2$ and $N=3$ shows that the model has an anisotropic
stable fixed point governing the continuous phase transitions with new
critical exponents. This is supported by  the estimate of the critical
dimensionality $N_c=1.445(20)$ obtained from six loops via the exact
relation $N_c=\frac{1}{2} n_c$ established for the complex and real
hypercubic models.
\end{abstract}

\vspace{.5cm}
\vspace{1cm}

\nid
~~\qquad Published in: \quad 
{\sl Physical Review} B {\bf 64}, 214423 (2001).

\vspace{5cm}

\rightline{{\sl Typeset using} \LaTeX}
\end{titlepage}

\section{Introduction}
\label{sec:1}

In this paper we investigate the critical properties of phase transitions in
certain antiferromagnets involving an increase of the unit cell
in one or more directions at the critical temperature. They are known to be
described by a generalized $2n$-component ($n \ge 2$)
Ginzburg-Landau model with three independent quartic terms
\be
\nn
H &=&
\int d^{~D}x \Bigl[{1 \over 2} \sum_{i=1}^{2n} (m_0^2~ \vp_i^2
 + \vec \nabla \vp_i \vec \nabla \vp_i)
 + {u_0 \over 4!} \Bigl(\sum_{i=1}^{2n} \vp_i^2 \Bigr)^2 \\
\nn \\
  &+& {v_0 \over 4!} \sum_{i=1}^{2n} \vp_i^4
 + {2 z_0 \over 4!} \sum_{i=1}^n \vp_{2i-1}^2\vp_{2i}^2 \Bigr]
\label{eq:Ham}
\ee
associated with isotropic, cubic, and tetragonal interactions, respectively
\cite{Muk}. Here $\vp_i$ is the real vector order parameter in $D=4 - \ve$
dimensions and $m_0^2$ is proportional to the deviation
from the mean-field transition point.
According to the universality hypothesis, the critical
phenomena in various
physical systems must not depend on microscopic details, but are determined
solely by the spatial dimensionality of the system, the interaction range
as well as the symmetry and the dimensionality of the order parameter field.
The anomalous behavior of the thermodynamic quantities and the correlation
radius $r_c$ are expressed  by power laws in the critical region.

When $n=2$, Hamiltonian (\ref{eq:Ham}) describes the antiferromagnetic
phase transitions in TbAu$_2$ and DyC$_2$ and the structural phase
transition in the NbO$_2$ crystal. The phase transitions in the
helical magnets Tb,
Dy, and Ho belong to the same class of universality \cite{Bak1}.
Another physically important case $n=3$ is relevant to the antiferromagnetic
phase transitions in such substances as K$_2$IrCl$_6$, TbD$_2$, and
Nd. The phase transition in antiferromagnet MnS$_2$ belongs to the same
class of universality \cite{Bak1}. All these phase transitions are predicted
to be of second order\footnote{An interesting type of multisublattice
antiferromagnets, such as MnO, CoO, FeO, and NiO, was studied in Ref.
\cite{Dz,Bak,Muk1}. It was shown, in the leading orders in $\ve$, that the
phase transitions in these substances are of first order.} \cite{Bak1,Muk1}.
This is confirmed by several experiments (see Ref. \cite{TMTB}
and references therein). However the experimental data were insufficiently
accurate to provide reliable values of critical exponents and the obtained
estimates \cite{Als,Pynn,Sch} were found to differ significantly from the
theoretically expected numbers. Such a distinction should be interpreted as
a manifestation of influence of various possible effects in the real
substances such as the presence of defects.

For the first time, the magnetic and structural phase transitions described
by model (\ref{eq:Ham}) were studied in the framework of the renormalization
group (RG) by Mukamel and Krinsky within the lowest
orders in $\ve$ \cite{Muk}. A three-dimensionally
stable fixed point (FP) with coordinates $u^* > 0$,
$v^* = z^* > 0$ was predicted\footnote{Following Mukamel \cite{Muk}, we call
this point "unique".}. That point was shown to determine
a new universality class with a specific set of critical exponents.
However, for the physically important case $n=2$, the critical exponents of
the unique point turned out to be exactly the same as those of
the $O(4)$-symmetric one.

For the years, an alternative analysis of critical behavior of the model, the
RG approach in three dimensions, was carried out within the two- and
three-loop approximations \cite{Shp,VS}. Those investigations gave the same
qualitative predictions: the unique stable FP does exist on the 3D
RG flow diagram. By using different resummation procedures, the
critical exponents computed at this point proved to be close to the
exponents
of the Bose FP ($u=0$, $v=z>0$) rather than the isotropic ($O(n)$-symmetric;
$u>0$, $v=z=0$) one. It was also shown that the unique and Bose FPs are
very close to each other, so that they may interchange their stability in the
next orders of RG approximation \cite{VS}.

Recently, the critical properties of the model were analyzed in third order in
$\ve$ \cite{BGMV,MV-1}. Investigation of the FP stability and calculation of
the critical dimensionality $n_c$ of the order parameter, separating two
different regimes of critical behavior\footnote{When $n > n_c$ the unique FP
is stable in 3D while for $n < n_c$ the stable FP is the isotropic one.},
confirmed that model (\ref{eq:Ham}) has the unique stable FP at $n=2$
and $n=3$. However, the twofold degeneracy of the stability matrix eigenvalues
at the one-loop level was observed for this FP \cite{MV-1}. That degeneracy
was shown to cause a substantial decrease of the accuracy expected within the
three-loop approximation and powers of $\sqrt{\ve}$ to appear in the
expansions\footnote{Similar phenomenon was observed earlier in studying the
impure Ising model (see Ref. \cite{Khm}). Half-integer powers in $\ve$
arising in that model have different origin but also lead to the loss of
accuracy.}.
So, computational difficulties were shown to grow faster than the
amount of essential information one may extract from high-loop approximations.
That resulted in the conclusion that the $\ve$-expansion method is
not quite effective for the given model.

Another problem associated with model (\ref{eq:Ham}) is the question whether
the unique FP is really stable in 3D, thus leading to a new class of
universal behavior, or its stability is only an effect of insufficient
accuracy of the RG approximations used. Indeed, there are general
nonperturbative theoretical arguments indicating that the only stable FP in
3D may be the Bose one and the phase transitions of interest should be
governed by that stable FP \cite{CB}. The point is that when $v=z$, the model
(\ref{eq:Ham}) describes $n$ interacting Bose systems. As was shown by Sak
\cite{Sak}, the interaction term can be represented as the product of the
energy operators of various two-component subsystems. It was also found that
one (the smallest) of the eigenvalue exponents characterizing the evolution of
this term under the renormalization group in a neighborhood of the Bose FP
is proportional to the specific heat exponent $\al$. Since $\al$ is believed
to be negative at that point, and that is supported experimentally \cite{Lipa}
and theoretically \cite{BNM78,GdZ98,Kl,CPRV}, the interaction is irrelevant.
Consequently, the Bose FP should be stable in three dimensions.
However, up to now, this conclusion found no confirmation within the RG
approach, so the model (\ref{eq:Ham}) poses a certain challenge to the
perturbative methods in the theory of critical phenomena.
It is therefore highly desirable to extend already known $\ve$ expansions
for the stability matrix eigenvalues, critical exponents and the critical
dimensionality in order to apply more sophisticated resummation techniques
to longer expansions.

Thus, in the present work we firstly avoid the problem of the eigenvalues
degeneracy in model (\ref{eq:Ham}) by analyzing the critical behavior
of the equivalent complex $N$-component order parameter model with
the effective Hamiltonian
\be
H =
\int d^D x \Bigl[{1 \over 2}\sum_{i=1}^{N}( m_0^2 \psi_i\psi^*_i
 + \vec {\nabla} \psi_i \vec {\nabla} \psi^*_i)
 + \frac{u_0}{4!}\ \sum_{i,j=1}^{N}\psi_i \psi^*_i \psi_j \psi^*_j
 + \frac{v_0}{4!}\ \sum_{i=1}^{N}\psi_i \psi_i \psi^*_i \psi^*_i \Bigr]
\label{eq:Ham1}
\ee
comprising the isotropic and cubic interactions\footnote{The model
with the complex vector order parameter was considered by
Dzyaloshinskii \cite{Dz77} in studying the phase transitions in
DyC$_2$, TbAu$_2$ ($N=2$) and TbD$_2$, MnS$_2$, and Nd
($N=3$).}. Note that this Hamiltonian also determines the real hypercubic
model \cite{Ah} relevant to the magnetic and structural phase transitions
in a cubic crystal if $\psi_i$ is thought to be the real $n$-component
order parameter. The model (\ref{eq:Ham1}) comes out exactly from model
(\ref{eq:Ham}) at $v_0=z_0$ and it is free from the eigenvalues degeneracy.
Secondly, we examine the existence of an anisotropic stable FP in model
(\ref{eq:Ham1}) on the basis of the higher-order $\ve$ expansion. Namely,
using dimensional regularization and the minimal subtraction scheme
\cite{Hooft}, we derive the four-loop RG functions as power series in $\ve$
and analyze the FP stability. For the first time, we give numerical estimates
for the stability matrix eigenvalues from $\ve$ expansions using the
Borel-Leroy transformation with a conformal mapping \cite{LgZ,VKT79}.
This allows us to carry out the careful analysis of the stability of all the
FPs of the model. We establish the exact relation $N_c=\frac{1}{2} n_c$
between the critical (marginal) spin dimensionalities of the complex and the 
real hypercubic models and obtain the
estimate $N_c=1.445(20)$ using six-loop results of Ref. \cite{CPV}.
We show that the anisotropic (complex cubic; $u\ne0$, $v\ne0$) stable FP
of model (\ref{eq:Ham1}), being a counterpart of the unique point in model
(\ref{eq:Ham}), does exist on 3D RG flow diagram at $N > N_c$. For this
stable FP we give more accurate critical exponents estimates in comparison
with the previous three-loop results \cite{MV-1} by applying the summation
technique of Ref. \cite{MV-2} to the longer series.

\section{Four-loop $\ve$-expansions, resummation
and the fixed point stability}
\label{sec:2}

To deduce RG expansions for the $\bt$-functions and critical exponents
one needs to calculate a set of Feynman graphs, each of them comprising
three factors: the combinatorial coefficient, the result of
tensor contractions and the integral value associated with the diagram.
The combinatorial factors and the values of integrals were found earlier in
Ref. \cite{VKT79}. To evaluate the tensor contractions for the vertex
and mass
diagrams, the tensors $G_1$ and $G_2$ corresponding to the
isotropic and cubic
interactions in Hamiltonian (\ref{eq:Ham1}) are introduced; they have a
simple symmetrized form:
\begin{eqnarray}
 G^{\>\>\>\al\bt\mu\nu}_{1\> ijkl}&=&\frac{1}{3}\Bigl(
\delta^{\al\bt}\delta^{\mu\nu} \delta_{ij}\delta_{kl}+
\delta^{\al\mu}\delta^{\bt\nu} \delta_{ik}\delta_{jl}+
\delta^{\al\nu}\delta^{\bt\mu} \delta_{il}\delta_{kj}\Bigr) \ ,
\nn \\
 G^{\>\>\>\al\bt\mu\nu}_{2\> ijkl}&=&
\delta^{\al\bt}\delta^{\al\mu}\delta^{\al\nu} \delta_{ij}
\delta_{ik}\delta_{il} \ ,
\label{eq:Tenz}
\nn
\end{eqnarray}
where $\{\al,\bt,\mu,\nu\} = 1, 2$ and $\{i,j,k,l\} = 1, \ldots, N$.

Further, normalizing conditions on the renormalized one-particle irreducible
inverse Green functions $\Gamma^{(2)}_R$ and vertices $\Gamma^{(4)}_R$
given by corresponding Feynman diagrams must be imposed. Within
the massless theory they are as follows
\begin{equation}
\begin{array}{lcrc}
\Gamma_R^{(2)}(\{p\};\mu, u, v)\Big\arrowvert_{p^2 = 0} &=& 0 & \ ,
\nonumber \\
{\partial \over {\partial p^2}} \Gamma_R^{(2)} (\{p\}; \mu, u, v)
\Big\arrowvert_{p^2 = \mu^2} &=& 1 & \ ,
\nonumber \\
\Gamma_{UR}^{(4)} (\{p\}; \mu, u, v)&=& \mu^{\ve} u & \ ,
\label{eq:NC1} \\
\Gamma_{VR}^{(4)} (\{p\}; \mu, u, v)&=& \mu^{\ve} v &
\nonumber
\end{array}
\end{equation}
with one more condition on the $|\psi|^2$ insertion
\begin{eqnarray}
\Gamma_R^{(1,2)}(\{p\},\{q\};\mu,u,v)
\Big\arrowvert_{{p^2 = q^2  = \mu^2 }\atop{pq=-{1\over3}}\mu^2} &=& 1
\ .
\label{eq:NC2}
\end{eqnarray}
Here $m$, $u$, and $v$ are the renormalized mass and
dimensionless coupling constants, $\mu$ is an arbitrary
mass parameter introduced for dimensional regularization.
The vertices $\Gamma_u^{(4)}$ and $\Gamma_v^{(4)}$ are connected with
the vertex function without external lines
$$
 \Gamma_{\>\>\>ijkl}^{(4)\>\>\>\al\bt\mu\nu}=
 \Gamma_u^{(4)} \cdot G^{\>\>\>\al\bt\mu\nu}_{1\> ijkl}+
 \Gamma_v^{(4)} \cdot G^{\>\>\>\al\bt\mu\nu}_{2\> ijkl} \ .
$$
>From renormalization conditions (\ref{eq:NC1}) and (\ref{eq:NC2}),
the expansions for the renormalization constants $Z_{\psi}$,
$Z_u$, $Z_v$, and $Z_{|\psi|^2}$ may be obtained.
These constants relate the bare mass $m_0$ and coupling
constants $u_0$, $v_0$ of the Hamiltonian (\ref{eq:Ham1}) to
the corresponding physical parameters:
\begin{equation}
   m_0^2 = \frac{Z_{|\psi|^2}}{Z_{\psi}} m^2=Z_m m^2 \ ,\quad
   u_0   = \mu^{\ve} \frac{Z_u}{Z^2_{\psi}} u \ , \quad
   v_0   = \mu^{\ve} \frac{Z_v}{Z^2_{\psi}} v \ . \quad
\label{eq:Conn}
\end{equation}
So, with relations (\ref{eq:Conn}) taken  into account, the
$\bt$-functions and critical exponents can be calculated via the
formulas
\begin{equation}
\begin{array}{lcr}
\frac{\partial \ln u_0}{\partial u} \bt_u +
\frac{\partial \ln u_0}{\partial v} \bt_v &=& - \ve \ ,
\nn \\
\nn \\
\frac{\partial \ln v_0}{\partial u} \bt_u +
\frac{\partial \ln v_0}{\partial v} \bt_v &=& - \ve \ ,
\label{eq:Bf}
\nn
\end{array}
\end{equation}
\begin{equation}
\begin{array}{lcr}
\eta(u,v)&=&
2 \frac{\partial \ln Z_{\psi}}{\partial u} \bt_u +
2 \frac{\partial \ln Z_{\psi}}{\partial v} \bt_v \ ,
\nn \\
\nn \\
\eta_2(u,v)&=& \>\>\>\>
2 \frac{\partial \ln Z_{|\psi|^2}}{\partial u} \bt_u +
2 \frac{\partial \ln Z_{|\psi|^2}}{\partial v} \bt_v
\label{eq:EtEt2}
\end{array}
\end{equation}
where $\bt_g \equiv \frac{\partial g}{\partial |\ln \mu|}$,
$g=\{u,v\}$. The critical exponents $\eta$ and $\eta_2$ are found
by substituting zeros of the $\bt$-functions into expressions
(\ref{eq:EtEt2}). The critical exponent $\gm$ is calculated through the
known scaling relation $\gm^{-1} = 1 + \frac {\eta_2}{2 - \eta}$.

The four-loop $\ve$ expansions for the $\bt$-functions of model
(\ref{eq:Ham1}) were recently obtained in Ref. \cite{MV00p}. They read
\be
\bt_u &=& \ve u -u^2 - \frac{4}{N + 4} u v
 + \frac{1}{(N+4)^2} \Bigl[3 u^3 (3 N + 7)
 + 44 u^2 v + 10 u v^2 \Bigr]
\nn \\
&-& \frac{1}{(N + 4)^3} \Bigl[\frac{u^4}{4} \Bigl(48 \zt(3) (5 N+11)
 +33 N^2+461 N+740 \Bigr)
\nn \\
&+& u^3 v \Bigl(384 \zt(3) +79 N + 659 \Bigr)
 + \frac{u^2 v^2}{2} \Bigl(288 \zt(3) +3 N +1078 \Bigr)
 + 141 u v^3 \Bigr]
\nn \\
&-& \frac{1}{(N + 4)^4} \Bigl[\frac{u^5}{12} \Bigl(- 48 \zt(3) (63 N^2
 + 382 N + 583) + 144  \zt(4)( 5 N^2
\nn \\
&+& 31 N +44) - 480 \zt(5) (4 N^2
 + 55 N + 93)+ 5N^3 - 3160 N^2
\nn \\
&-& 20114 N- 24581 \Bigr)
 - \frac{2 u^4 v}{3} \Bigl(12 \zt(3) (3 N^2
 + 276 N + 1214) - 36 \zt(4)
\nn \\
&\times& (19 N + 85) + \zt(5) (2400 N + 23040) - 28 N^2
 + 3957 N +15967 \Bigr)
\nn \\
&-& \frac{u^3 v^2}{3} \Bigl(72 \zt(3) (19 N + 426) - 4032 \zt(4)
 + 39840 \zt(5) + 1302 N + 46447 \Bigr)
\nn \\
&+& \frac{2 u^2 v^3}{3} \Bigl(60 \zt(3) (N- 84) - 792 \zt(4)
 - 4800 \zt(5) - 125 N - 12809 \Bigl)
\nn \\
&-& \frac{u v^4}{2} \Bigl(400 \zt(3) + 768 \zt(4) + 3851 \Bigr) \Bigr] \ ,
\label{eq:Bu}
\ee
\be
\bt_v &=& \ve v - \frac{1}{N+4}(6 u v + 5 v^2) + \frac{1}{(N+4)^2}
\Bigl[u^2 v (5 N + 41) + 80 u v^2 + 30 v^3 \Bigr]
\nn \\
&-& \frac{1}{(N + 4)^3} \Bigl[\frac{u^3 v}{2} \Bigl(96 \zt(3) (N + 7)
 - 13 N^2 + 184 N + 821 \Bigr)
\nn \\
&+& \frac{u^2 v^2}{4} \Bigl(4032 \zt(3) + 59 N
 + 5183 \Bigr) + u v^3 \Bigl(768 \zt(3) + 1093 \Bigr)
 + \frac{v^4}{2} \Bigl(384 \zt(3) + 617 \Bigr) \Bigr]
\nn \\
&-& \frac{1}{(N + 4)^4} \Bigl[\frac{u^4 v}{4} \Bigl(48 \zt(3) (N^3
 -12 N^2 - 140 N - 567) + 144 \zt(4) (2 N^2
\nn \\
&+& 17 N +45) - 3360 \zt(5) (3 N + 13)
 - 29 N^3 - 28 N^2 - 6958 N - 19679 \Bigr)
\nn \\
&+& \frac{u^3 v^2}{3} \Bigl(12 \zt(3) (9 N^2 - 591 N -7028)
 + \zt(4) (3528 N + 21240) - 480 \zt(5)
\nn \\
&\times& (10 N + 287) + 61 N^2 - 5173 N - 66764 \Bigr)
 - \frac{u^2 v^3}{3} \Bigl(1800 \zt(3) (N + 62)
\nn \\
&-& 144 \zt(4) (8 N + 203) + 172800 \zt(5) + 56 N + 93701 \Bigr)
 - 4 u v^4 \Bigl(5090 \zt(3)
\nn \\
&-& 1296 \zt(4) + 7600 \zt(5) + 4503 \Bigr)
 + \frac{v^5}{2} \Bigl(- 8224 \zt(3) + 1920\zt(4)
\nn \\
&-& 12160 \zt(5)-7975 \Bigr) \Bigr]
\label{eq:Bv}
\ee

\nid
where $\zt(3)$, $\zt(4)$, and $\zt(5)$ are the
Riemann $\zt$ functions.
For the critical exponents we find
\be
\eta&=&
\frac{1}{(N+4)^2} \Bigl[u^2 (N+1) + 4 u v + 2 v^2 \Bigr]
\nn \\
&-& \frac{1}{(N+4)^3} \Bigl[\frac{u^3}{2} (N+1) (N+4)
 + 3 u^2 v (N+4) + 15 u v^2 + 5 v^3 \Bigr]
\nn \\
&+& \frac{1}{(N+4)^4} \Bigl[- \frac{5}{4} u^4 (N+1) \Bigl(N^2
 - 9 N - 25 \Bigr) - 10 u^3 v \Bigl(N^2
\nn \\
&-& 9 N - 25\Bigr) + 495 u^2 v^2 + 330 u v^3 + \frac{165}{2} v^4 \Bigr] \ ,
\label{eq:ETAuv}
\ee
\be
\eta_2&=&
- \frac{1}{N+4} \Bigl[2 u (N+1) + 4 v \Bigr]
+ \frac{1}{(N+4)^2} \Bigl[6 u^2 (N+1) + 24 u v + 12 v^2 \Bigr]
\nn \\
&-& \frac{1}{(N+4)^3} \Bigl[\frac{u^3}{2} \Bigl(31 N^2
+ 146 N + 115 \Bigr) + 3 u^2 v (31 N+115)
\nn \\
&+& 3 u v^2 (N+145) + 146 v^3\Bigr]
+ \frac{2}{(N+4)^4} \Bigl[\frac{u^4}{6} \Bigl(12\zt(3) (3 N^3
+ 8 N^2
\nn \\
&+& 22 N + 17) + 36\zt(4) (5 N^2 + 16 N + 11)
- 4 N^3 + 977 N^2
\nn \\
&+& 3016 N + 2035 \Bigr)
+ \frac{4 u^3 v}{3} \Bigl(12 \zt(3) (3 N^2 + 5 N + 17)
+ \zt(4) (180 N
\nn \\
&+& 396) - 4 N^2 + 981 N + 2035 \Bigr)
+ 4 u^2 v^2 \Bigl(\zt(3) (18 N+132)
\nn \\
&+& \zt(4) (18 N+270) + 115 N + 1391 \Bigr)
- \frac{2 u v^3}{3} \Bigl(60 \zt(3) (N-11)
\nn \\
&-& 1152 \zt(4) - 125 N - 5899 \Bigr)
+ 4 v^4 \Bigl(25 \zt(3) + 48 \zt(4) + 251 \Bigr) \Bigr] \ ,
\label{eq:ETA2uv}
\ee
\be
\gm^{-1}&=&
1- \frac{1}{(N+4)} \Bigl[u (N+1) + 2 v \Bigr]
+ \frac{3}{(N+4)^2} \Bigl[u^2 (N+1) + 4 u v + 2 v^2 \Bigr]
\nn \\
&-& \frac{1}{4 (N+4)^3} \Bigl[3 u^3 (N+1) (11 N+39)
 + 18 u^2 v (11 N+39)
\nn \\
&+& 10 u v^2 (N+89) + 300 v^3 \Bigr]
+ \frac{1}{12 (N+4)^4} \Bigl[u^4 \Bigl(24\zt(3) (N+1) (3 N^2
\nn \\
&+& 5 N + 17) + 72\zt(4) (N+1) (5 N+11) - 5 (N+1) (N^2 - 399 N
\nn \\
&-& 820) \Bigr) + 8 u^3 v \Bigl(24\zt(3) (3 N^2 + 5 N + 17)
+ \zt(4) (360 N+792)
\nn \\
&-& 5 (N^2 - 399 N - 820) \Bigr) + 6 u^2 v^2 \Bigl(48\zt(3) (3 N+22)
\nn \\
&+& 144\zt(4) (N+15) + 953 N + 11227 \Bigl)
- 2 u v^3 \Bigl(240\zt(3) (N-11)
\nn \\
&-& 4608\zt(4) - 515 N - 23845 \Bigl) + 12 v^4 \Bigl(100\zt(3)
+ 192\zt(4) + 1015 \Bigl) \Bigr] \ .
\label{eq:GMRuv}
\ee

>From the system of equations $\bt_u(u^*, v^*)=0$, $\bt_v(u^*, v^*)=0$, one can
calculate all the FPs of model (\ref{eq:Ham1}), which are given by the power
series in $\ve$:
$$u^*=u^*(\ve)=\sum^\infty_{k=1}u_k \ve^k \ , \qquad
v^*=v^*(\ve)=\sum^\infty_{k=1}v_k \ve^k \ .$$
There exist four FPs, one of them (Gaussian) is trivial:
\vspace{0.5cm}

\begin{tabular}{llll}
{\bf 1.}&\multicolumn{3}{l}{\bf  Gaussian FP}
\\[6pt]
&$u^*$&=&$v^*\> \> = \> \> 0$ .
\\[6pt]&&&
\end{tabular}

\begin{tabular}{llll}
{\bf 2.}&\multicolumn{3}{l}{\bf Isotropic or $O(N)$-symmetric FP}
\\[6pt]
&$u^*$&=&$ \ve+\frac{3}{(N+4)^2} (3 N+7) \ve^2
  - \frac{1}{4(N+4)^4} \Bigl(48 \zt(3) (N+4)(5N+11)$
\\[6pt]&
&$+$& $33 N^3 - 55 N^2 - 440N-568 \Bigr) \ve^3$
\\[6pt]&
&$+$&$ \frac{1}{12(N+4)^6} \Bigl(48\zt(3)(N+4) \Bigl(63 N^3-41 N^2
  -949N-1133 \Bigr)$
\\[6pt]&
&$-$&$ 144\zt(4)(N+4)^3(5N+11)+480 \zt(5)(N+4)^2 \Bigl(4 N^2$
\\[6pt]&
&$+$&$ 55 N+93 \Bigr) - 5 N^5 -1335 N^4 -2(698 N^3
  - 3299 N^2$
\\[6pt]&
&$-$&$ 9666 N-8278 \Bigr) \ve^4$ ,
\\[6pt]
&$v^*$&=&$\>0$ .
\\[6pt]&&&
\end{tabular}

\begin{tabular}{llll}
{\bf 3.}&\multicolumn{3}{l}{\bf Bose FP}
\\[6pt]
&$u^*$&=&$\>0$ ,
\\[6pt]
&$v^*$&=&$\frac{1}{5}(N+4) \ve + \frac{6}{25}(N+4)\ve^2
+ \frac{1}{1250}(N+4) \Bigl(103-384 \zt(3) \Bigr) \ve^3$
\\[6pt]&
&$+$&$ \frac{1}{6250}(N+4) \Bigl(-3296\zt(3)-1920 \zt(4)
  +12160 \zt(5)+265 \Bigr) \ve^4$ .
\\[6pt]&&&
\end{tabular}

\begin{tabular}{llll}
{\bf 4.}&\multicolumn{3}{l}{\bf Complex cubic FP}
\\[6pt]
&$u^*$&=&$\frac{1}{5N-4}(N+4) \ve
  + \frac{1}{(4-5N)^3}(N+4) \Bigl(70 N^2 -205 N+139 \Bigr) \ve^2$
\\[6pt]&
&$+$&$\frac{1}{4(5N-4)^5}(N+4) \Bigl(48 \zt(3)(5N-4) \Bigl(64 N^3
  - 188 N^2 +151 N$
\\[6pt]&
&$-$&$ 23 \Bigr)-6370 N^4-24149 N^3
  +144719 N^2-197208 N+83256 \Bigr) \ve^3$
\\[6pt]&
&$+$&$ \frac{1}{12 (5N-4)^7}(N+4) \Bigl(48\zt(3) \Bigl(22200 N^6
  +110580 N^5-754767 N^4$
\\[6pt]&
&$+$&$ 1326821 N^3 -887361 N^2+132711 N+49508 \Bigr)
  +144\zt(4) (5N-4)^3$
\\[6pt]&
&$\times$&$ \Bigl(96 N^3-276 N^2+221 N-37 \Bigr) -
480\zt(5) \Bigl(15200 N^6$
\\[6pt]&
&$-$&$ 79920 N^5+168488 N^4-183439 N^3
+109827 N^2 -34792 N$
\\[6pt]&
&$+$&$ 4656 \Bigr) -99250 N^6 +2692575 N^5 -36725295 N^4
+140337844 N^3$
\\[6pt]&
&$-$&$ 230649570 N^2 +174742836 N -50310868 \Bigr) \ve^4$ ,
\\[6pt]
\end{tabular}

\begin{tabular}{llll}
&$v^*$&=&$ \frac{1}{5N-4}(N+4)(N-2) \ve
  + \frac{1}{(5N-4)^3}(N+4) \Bigl(30 N^3+25 N^2$
\\[6pt]&
&$-$&$ 217 N+166 \Bigr) \ve^2
  + \frac{1}{4(4-5N)^5}(N+4) \Bigl(96 \zt(3)(5N-4) \Bigl(8 N^4
  +16 N^3$
\\[6pt]&
&$-$&$ 88 N^2 +75 N-9 \Bigr)
  -1030 N^5-2751 N^4-46033 N^3$
\\[6pt]&
&$+$&$ 2 \Bigl(103795 N^2-133668 N+54904 \Bigr) \Bigr) \ve^3$
\\[6pt]&
&$+$&$ \frac{1}{12 (4-5N)^7}(N+4) \Bigl(48\zt(3) \Bigl(10300 N^7
  - 39580 N^6+328467 N^5$
\\[6pt]&
&$-$&$ 1208826 N^4 +1806293 N^3 -1074204 N^2
  +94274 N +82968 \Bigr)$
\\[6pt]&
&$+$&$ 144 \zt(4) \Bigl(2000 N^7+5200 N^6-58660 N^5+142451 N^4$
\\[6pt]&
&$-$&$ 160415 N^3 + 92612 N^2-25680 N+2496 \Bigr)
  -480\zt(5) \Bigl(3800 N^7$
\\[6pt]&
&$-$&$2280 N^6 -47048 N^5+134947 N^4-162421 N^3
+101278 N^2$
\\[6pt]&
&$-$&$ 32704 N +4448 \Bigr)-39750 N^7+1242425 N^6-3090975 N^5$
\\[6pt]&
&$-$&$ 35240910 N^4 +171686590 N^3 -295865304 N^2$
\\[6pt]&
&$+$&$ 226373292 N - 65077096 \Bigr) \ve^4$ .
\\[6pt]&&&
\end{tabular}

The stability properties of the FPs are controlled by the
eigenvalue exponents $\om_i$ of the stability matrix
\begin{equation}
\Omega  \> = \>
\left(
\begin{array}{cc}
    \frac{\partial\bt_u(u,v)}{\partial u} &
    \frac{\partial\bt_u(u,v)}{\partial v} \\
    \frac{\partial\bt_v(u,v)}{\partial u} &
    \frac{\partial\bt_v(u,v)}{\partial v}
\end{array}
\right)
\label{eq:Om}
\end{equation}
taken at the given FPs. We calculated the eigenvalue exponents at all the FPs.
For the most intriguing Bose and complex cubic FPs they are \cite{MV-jl}
\be
\nn
\om_1 &=& - \frac{1}{2} \ve + \frac {6}{20} \ve^2
+ \frac{1}{8} \Bigl[- \frac {257}{125}
- \frac {384}{125} \zt(3) \Bigr] \ve^3 \\
\nn
      &+& \frac{1}{16} \Bigl[\frac {5109}{1250}
+ \frac {624}{125} \zt(3)
- \frac {576}{125} \zt(4)
+ \frac {3648}{125} \zt(5) \Bigr] \ve^4  \ ,
\ee
\be
\nn
\om_2 &=& \frac {1}{10} \ve - \frac {14}{100} \ve^2
+ \frac{1}{8} \Bigl[- \frac {311}{625}
+ \frac {768}{625} \zt(3) \Bigr] \ve^3 \\
      &+& \frac{1}{16} \Bigl[- \frac {61}{250}
+ \frac {3752}{3125} \zt(3)
+ \frac {1152}{625} \zt(4)
- \frac {4864}{625} \zt(5) \Bigr] \ve^4
\label{eq:Bose}
\ee
at the Bose FP and
\be
\nn
\om_1 &=& - \frac{1}{2} \ve
+ \frac { \Bigl(60 N^3 - 160 N^2
+ 181 N - 85 \Bigr)}{4 (5 N - 4)^2 (2 N - 1)} \ve^2 \\
\nn
&-& \frac {1}{16 (2 N - 1)^3 (5N - 4)^4}
\Bigl[\Bigl(30720 N^7 - 178176 N^6 + 456960 N^5 \\
\nn
&-& 648384 N^4 + 534336 N^3 - 251952 N^2
+ 62640 N - 6336 \Bigr) \zt(3) \\
\nn
&+& 20560 N^7 - 165328 N^6 + 644392 N^5 - 1406864 N^4 \\
\nn
&+& 1756745 N^3 - 1224341 N^2 + 433704 N - 59052 \Bigr] \ve^3 \\
\nn
&+& \frac {1}{64 (2 N - 1)^5 (5 N - 4)^6}
\Bigl[ \Bigl(9984000 N^{11} - 149237760 N^{10} \\
\nn
&+& 921189888 N^9 - 3096268032 N^8 + 6362845824 N^7 \\
\nn
&-& 8473037952 N^6 + 7511188512 N^5 - 4452728592 N^4 \\
\nn
&+& 1734564864 N^3 - 423296208 N^2 + 58187520 N \\
\nn
&-& 3401472 \Bigr) \zt(3)
+ \Bigl(- 9216000 N^{11} + 77414400 N^{10} \\
\nn
&-& 299013120 N^9 + 693626112 N^8 - 1064454912 N^7 \\
\nn
&+& 1127834496 N^6 - 838693440 N^5 + 436803408 N^4 \\
\nn
&-& 155991312 N^3 + 36370368 N^2 - 4983552 N \\
\nn
&+& 304128 \Bigr) \zt(4) + \Bigl(58368000 N^{11} - 501964800 N^{10} \\
\nn
&+& 1976709120 N^9 - 4653603840 N^8 + 7224314880 N^7 \\
\nn
&-& 7735272960 N^6 + 5820821760 N^5 - 3078575040 N^4 \\
\nn
&+& 1122437280 N^3 - 268990560 N^2 + 38181120 N \\
\nn
&-& 2434560 \Bigr) \zt(5) + 8174400 N^{11} - 111638720 N^{10} \\
\nn
&+& 782999600 N^9 - 3263360352 N^8 + 8565392076 N^7 \\
\nn
&-& 14703626172 N^6+ 16826740917 N^5 - 12839720112 N^4 \\
\nn
&+& 6396391932 N^3 - 1976965857 N^2 + 340410916 N \\
\nn
&-& 24804436 \Bigr] \ve^4 \ ,
\ee
\be
\nn
\om_2 &=& - \frac {(N - 2)}{2 (5 N - 4)} \ve
+ \frac {(N - 1) \Bigl(140 N^3 - 600 N^2 + 567 N - 70 \Bigr)}
{4 (2 N - 1) (5 N - 4)^3} \ve^2  \\
\nn
&-&  \frac {(N - 1)} {16 (2 N - 1)^3 (5 N - 4)^5}
\Bigl[ \Bigl(61440 N^7 - 402432 N^6 + 975744 N^5 \\
\nn
&-& 1193856 N^4 + 827808 N^{3} - 337536 N^2
+ 77424 N - 7872 \Bigr) \zt(3) \\
\nn
&-& 24880 N^7 + 140544 N^6 - 550920 N^5 + 1554928 N^4 \\
\nn
&-& 2500155 N^3 + 2090850 N^2 - 828628 N + 118872 \Bigr] \ve^3 \\
\nn
&-& \frac {(N - 1)} {64 (2 N - 1)^5 (5 N - 4)^7}
\Bigl[ \Bigl(12006400 N^{11} - 89251840 N^{10} \\
\nn
&+& 476114176 N^9 - 1821753600 N^8 + 4343448768 N^7 \\
\nn
&-& 6373076352 N^6 + 5838538224 N^5 - 3325387152 N^4 \\
\nn
&+& 1126329168 N^3 - 198020768 N^2 + 10046816 N \\
\nn
&+& 986752 \Bigr) \zt(3) + \Bigl(18432000 N^{11} - 168652800 N^{10} \\
\nn
&+& 652515840 N^9 - 1439023104 N^8 + 2036938752 N^7 \\
\nn
&-& 1962510336 N^6 + 1324196640 N^5 - 630472896 N^4 \\
\nn
&+& 209021904 N^3 - 46238400 N^2 + 6172416 N \\
\nn
&-& 377856 \Bigr) \zt(4) + \Bigl(- 77824000 N^{11}
+ 679014400 N^{10} \\
\nn
&-& 2474536960 N^9 + 5043287040 N^8 - 6405066240 N^7 \\
\nn
&+& 5277569280 N^6 - 2800588800 N^5 + 883784640 N^4 \\
\nn
&-& 113276640 N^3 - 20895040 N^2 + 9518080 N \\
\nn
&-& 1008640 \Bigr) \zt(5) - 2440000 N^{11} - 83641600 N^{10} \\
\nn
&+& 1064384720 N^9 - 5509282032 N^8 + 16208366916 N^7 \\
\nn
&-& 29918087352 N^6 + 35918140899 N^5 - 28188817515 N^4 \\
\nn
&+& 14172784689 N^3 - 4333185934 N^2 + 721222916 N \\
&-& 49452712 \Bigr] \ve^4
\label{eq:CC}
\ee
at the complex cubic one.

It is known that RG series are at best asymptotic, the coefficients
of the series $\sum_k f_k g^k $ behave at large $k$ as $k! k^{b} (- a)^k$
\cite{BNM78,Lip77,BLGZ77}.
An appropriate resummation procedure has to be applied, in order to extract
reliable physical information from them. To obtain the eigenvalues
estimates, we have used an approach based on the Borel-Leroy transformation
\be
 F(\ve;a,b) =\sum_{k=0}^\infty A_k(\la) \int_0^\infty
 e^{-\frac{x}{a \ve}} \Bigl(\frac{x}{a \ve}\Bigr)^b
 d\Bigl(\frac{x}{a \ve}\Bigr)
 \frac{z^k(x)}{[1-z(x)]^{2\la}}
 \label{eq:Bor}
\ee
modified with the conformal mapping $z=\frac{\sqrt{x+1}-1}{\sqrt{x+1}+1}$
\cite{LgZ,VKT79}, which does not require the knowledge of the exact
asymptotic high-order behavior of the series \cite{MV-2}.
The coefficients $A_k(\la)$ are determined from the equality
$B(x(z))=\frac{A(\la,z)}{(1-z)^{2\la}}$ where
the Borel-Leroy transform $B(x)$ is the analytical continuation
of the series $\sum_k \frac{f_k}{a^k\Gamma(b+k+1)} x^k$ absolutely
convergent in the unit circle, $f_k$ are the coefficients of the
original series. The numbers $a$ and $b$ characterize the main
divergent part of the series. Since, in practice, we deal with a piece
of the series only where the asymptotic regime might not be established,
we vary parameters $a$ and $b$ in a neighborhood of their exact values.
Our principle observation is that the result of processing $F(\ve;a,b)$ 
exhibits very weak dependence on the transformation parameters $a$ 
and $b$ varying
in a wide range. Thus, we put the stability of the result of
processing with respect to variation of $a$ and $b$ into the foundation
of our approach to resummation of divergent series\footnote{The exact 
values of the asymptotic parameters $a$ and $b$ were calculated for the
$O(n)$-symmetric models \cite{BLGZ77}, for the real cubic model \cite{CPV},
and recently for a number of more complicated models \cite{PV-z}.
Calculations show that the resummation method employed in this paper gives
the same results as the conventional technique using  the exact
values of the asymptotic parameters (see Refs. \cite{CPV} and \cite{MV-2})}
\cite{MV-2}. The parameter $\la$ is chosen from the condition of the most 
rapid convergence of the series
(\ref{eq:Bor}), that is from minimizing the quantity
$|1-\frac{F_l(\ve;a,b)}{F_{l-1}(\ve;a,b)}|$, where $l$ is the step of
truncation and $F_l(\ve;a,b)$ is the $l$-partial sum for $F(\ve;a,b)$.
If both eigenvalues are
negative, the FP is infrared stable and the critical
behavior of experimental systems undergoing second-order phase transitions
is determined only by that stable point. For the Bose and
the complex cubic FPs, our numerical results are displayed in Table I.

\begin{table}
\caption{Eigenvalue estimates obtained for the Bose (BFP)
and the complex cubic (CCFP) fixed points at $N=2$ and $N=3$ in the
four-loop order in $\ve$ ($\ve=1$) using the Borel-Leroy transformation with
a conformal mapping.}
\label{TabI}
\vspace{0.5cm}
\hspace{2cm}
\begin{tabular}{|c|l|l|l|l|}\hline
Type                &\multicolumn{2}{|c|}{$N=2$}&
                     \multicolumn{2}{|c|}{$N=3$}
                   \\[0pt]    \cline{2-5}
of FP               &\multicolumn{1}{|c|}{$\om_1$}
                    &\multicolumn{1}{|c|}{$\om_2$}
                    &\multicolumn{1}{|c|}{$\om_1$}
                    &\multicolumn{1}{|c|}{$\om_2$}
                   \\[0pt] \hline
BFP                 &$-0.395(25)$ & $0.004(5)$
                    &$-0.395(25)$ & $0.004(5)$
                   \\[0pt] \hline
CCFP                &$-0.392(30)$ & $-0.029(20)$
                    &$-0.400(30)$ & $-0.015(6)$
                   \\[0pt]\hline
\end{tabular}
\end{table}

\nid
It is seen that the complex cubic FP is absolutely stable in $D=3$
($\ve=1$), while the Bose point appears to be of the "saddle"
type\footnote{The fixed point is called to be of  the "saddle" type provided
their eigenvalue exponents $\om_1$ and $\om_2$ are of
opposite sings at the ($u, v$) plane.}.
However $\om_2$'s of either points are very small at the four-loop level,
thus implying that these points may swap their stability in the next order
of RG approximation. We can compare $\om_2$ at the
complex cubic FP quoted in Table I with the three-loop results of
Ref. \cite{Shp} obtained in the framework of RG approach directly
in 3D. Those estimates $\om_2=-0.010$ for $N=2$ and $\om_2=-0.011$
for $N=3$ are solidly consistent with ours.

In addition to the eigenvalues, we have computed the critical
dimensionality of the order
parameter of model (\ref{eq:Ham1}). The critical spin dimensionality $N_c$
for the complex cubic model is determined as that value of $N$, at
which the complex cubic fixed point coincides with the isotropic one.
Equivalently, for $N=N_c$ the second eigenvalue of the stability
matrix $\Omega$ taking at the complex cubic FP vanishes, $\om_2 = 0$. 
The four-loop $\ve$ expansion reads
$$N_c = 2 - \ve + \frac{5}{24} \Bigl[6 \zt(3) -1 \Bigr] \ve^2
 + \frac{1}{144} \Bigl[45 \zt(3)
 + 135 \zt(4) - 600 \zt(5) -1 \Bigr] \ve^3 .$$
Instead of processing this expression numerically, we have established the
exact relation $N_c=\frac{1}{2} n_c$ between the critical spin
dimensionalities of the complex and the real hypercubic models, which is
independent on the order of approximation used.
In fact, both models, the complex and the real cubic ones, exhibit
effectively the isotropic critical behavior at $N=N_c$ and $n=n_c$,
respectively. Therefore, because the complex $O(2N)$-symmetric
model is equivalent to the real $O(n)$-symmetric one, the relation
$2 N_c=n_c$ holds true. For $N>N_c$ the complex cubic FP
of model (\ref{eq:Ham1}) should be stable in 3D.

The five-loop $\ve$ expansion for $n_c$ was recently obtained in Ref.
\cite{KSf}.
Resummation of that series gave the estimate $n_c=2.894(40)$ (see Ref.
\cite{VKB}).
Therefore we conclude that $N_c=1.447(20)$ from the five-loops.
Practically the same estimate $N_c=1.435(25)$ follows from a
constrained analysis of $n_c$ taking into account $n_c=2$ in two
dimensions \cite{CPV}. From the recent pseudo-$\ve$ expansion
analysis of the real hypercubic model \cite{FHY} one can extract
$N_c=1.431(3)$. However the most accurate estimate $N_c=1.445(20)$ results
from the value $n_c=2.89(4)$ obtained on the basis of the numerical
analysis of the four-loop \cite{VKB} and the six-loop
\cite{CPV} 3D RG expansions for the $\bt$-functions of the real hypercubic
model. Since $N_c < 2$, the critical thermodynamics in the NbO$_2$ crystal,
in the antiferromagnets TbAu$_2$, DyC$_2$, K$_2$IrCl$_6$, TbD$_2$, MnS$_2$,
and Nd as well as in the gelical magnets Tb, Dy, and Ho should be controlled
by the complex cubic fixed point with a specific set of critical exponents,
in the frame of the given approximation.

\section{Critical Exponents and Conclusion}
\label{sec:3}

In the previous section we have shown that one of the four FPs is
absolute
stable in 3D. This is the anisotropic complex cubic FP. Now let us calculate
the critical exponents. To do this, substitute the coordinates of the FPs
found in Sec. II into the expressions for $\eta$ and $\gm^{-1}$
[see Eqs. (\ref{eq:ETAuv}) and (\ref{eq:GMRuv})]. For the stable complex 
cubic FP, this yields 
\be
\nn
\eta &=&
     \frac{\ve^2}{4 (5 N - 4)^2}(N - 1)(2 N - 1) +
     \frac{\ve^3}{16 (5 N - 4)^4}(N - 1) \Bigl[190 N^3 \\
\nn
     &-& 535 N^2 + 652 N - 324 \Bigr] -
     \frac{\ve^4}{64 (5 N - 4)^6} (N - 1)
     \Bigl[96 \zt(3) \Bigl(160 N^5 \\
\nn
     &-& 768 N^4 + 1572 N^3 - 1613 N^2 + 777 N - 132 \Bigr)
     - 10570 N^5 \\
     &+& 22691 N^4 + 68527 N^3 - 280399 N^2 + 326888 N
     - 127676 \Bigr] ,
\label{eq:ETA}
\ee
\be
\nn
\gm^{-1}
&=& 1 - \frac{3 \> \ve} {2 (5 N - 4)} (N - 1)
 + \frac{\ve^2}{4 (5 N - 4)^3} (N - 1)
   \Bigl[40 N^2 - 214 N \\
\nn
&+& 205 \Bigr]
- \frac{\ve^3}{8 (5 N - 4)^5} (N-1)
 \Bigl[12 \zt(3) (5 N - 4) \Bigl(32 N^3
 - 156 N^2 \\
\nn
&+& 159 N - 13 \Bigr) - 940 N^4 - 6748 N^3
+ 42681 N^2 - 67102 N \\
\nn
&+& 32558 \Bigr]
- \frac{\ve^4}{64 (5 N - 4)^7} (N-1) \Bigl[8 \zt(3) (5 N - 4)
\Bigl(5650 N^5 \\
\nn
&+& 58655 N^4 - 303683 N^3 + 396967 N^2
- 96502 N - 65510 \Bigr) \\
\nn
&+& 72 \zt(4) (5 N - 4)^3 \Bigl(32 N^3 - 156 N^2
+ 159 N - 13 \Bigr) \\
\nn
&-& 160 \zt(5) (5 N - 4)^2 \Bigl(304 N^4
- 1616 N^3 + 2424 N^2 - 1277 N \\
\nn
&+& 265\Bigr) - 54300 N^6 - 17220 N^5 - 7422827 N^4
+ 42427564 N^3 \\
&-& 86521137 N^2 + 76563448 N - 25005620 \Bigr] .
\label{eq:GMr}
\ee
The other critical exponents can be found through the known scaling relations.
The numerical estimates obtained via transformation (\ref{eq:Bor})
are collected in Table II. The critical exponents for the isotropic
and Bose FPs are also presented, for comparison. We can compare our
results with the available experimental data. For example, in the case of
the structural transition in NbO$_2$ crystal, the critical exponent of
spontaneous polarization was measured in Ref. \cite{Pynn}, $0.33< \bt < 0.44$.
Our estimate $\bt=0.371$ obtained using the data of Table II and scaling
relations lies in that interval.

\begin{table}
\caption{Critical exponents for the isotropic (IFP), the Bose
(BFP), and the complex cubic (CCFP) FPs at $N=2$ and $N=3$
calculated in the four-loop order in $\ve$ ($\ve=1$) using the Borel-Leroy
transformation with a conformal mapping.}
\label{TabII}
\vspace{0.5cm}
\hspace{0.75cm}
\begin{tabular}{|c|l|l|l|l|l|l|}\hline
Type                &\multicolumn{3}{|c|}{$N=2$}&
                     \multicolumn{3}{|c|}{$N=3$}
                   \\[0pt]    \cline{2-7}
of FP               &\multicolumn{1}{|c|}{$\eta$}
                    &\multicolumn{1}{|c|}{$\nu$}
                    &\multicolumn{1}{|c|}{$\gm$}
                    &\multicolumn{1}{|c|}{$\eta$}
                    &\multicolumn{1}{|c|}{$\nu$}
                    &\multicolumn{1}{|c|}{$\gm$}
                   \\[0pt]  \hline
IFP                 &$0.0343(20)$ & $0.725(15)$ & $1.429(20)$
                    &$0.0317(10)$ & $0.775(15)$ & $1.524(25)$
                   \\[0pt] \hline
BFP                 &$0.0348(10)$ & $0.664(7)$ & $1.309(10)$
                    &$0.0348(10)$ & $0.664(7)$ & $1.309(10)$
                   \\[0pt] \hline
CCFP                &$0.0343(20)$ & $0.715(10)$ & $1.404(25)$
                    &$0.0345(15)$ & $0.702(10)$ & $1.390(25)$
                   \\[0pt]\hline
\end{tabular}
\end{table}

In conclusion, the four-loop $\ve$-expansion analysis of the
Ginzburg-Landau model with cubic anisotropy and complex vector order parameter
relevant to the phase transitions in certain antiferromagnets with
complicated ordering has been carried out.
Investigation of the global structure of RG flows for the physically
significant cases $N=2$ and $N=3$ has shown that the anisotropic complex
cubic FP is absolutely stable in 3D. Therefore the critical
thermodynamics of the phase transitions of concern is governed by
this stable point with specific critical exponents. The critical dimensionality
$N_c=1.445(20)$ obtained from six loops supports this conclusion.
At the complex cubic FP, the critical exponents were calculated
using the Borel-Leroy resummation technique in combination with a conformal
mapping. For the structural phase transition in NbO$_2$, for the
antiferromagnetic phase transitions in TbAu$_2$ and DyC$_2$ as well as for
the phase transitions in the rare-earth metals Ho, Dy, and Tb, they were
shown to be close to the critical exponents of the $O(4)$-symmetric model.
In contrast to this, the critical exponents for the antiferromagnetic phase
transitions in K$_2$IrCl$_6$, TbD$_2$, MnS$_2$, and Nd turned out to be close
to the Bose ones.

Although our calculations show that the complex cubic FP, rather than 
the Bose one, is stable at the four-loop level, the eigenvalues $|\om_2|$ of 
both points are very small.
Therefore the situation is very close to marginal, and the FPs might
change their stability to opposite in the next order of
perturbation theory, so that the Bose point would occur stable.
This conjecture is in agreement with the recent six-loop RG study of
three-coupling-constant model (\ref{eq:Ham}) directly in three dimensions
\cite{PV-z}.
There, the authors argue the global stability of the Bose FP, 
although
the numerical estimate  $\om_2=-0.007(8)$ of the smallest stability matrix
eigenvalue at the Bose FP appears to be very small and
the apparent accuracy of the analysis does not exclude the opposite sign 
for $\om_2$. In this situation, it would be highly desirable to compare
the critical exponents values obtained theoretically with
values that could be determined from experiments, in order to verdict
which of the two FPs is really stable in physical space.
Finally, it would be also useful to investigate certain universal amplitude
ratios of the model because they vary much more among different universality
classes than exponents do and might be more effective as a diagnostic tool.

\section*{Acknowledgments}

We are grateful to Professor M. Henkel for helpful remarks. We
thank Dr. E. Vicari and Dr. E. Blagoeva for sending to us a copy
of their articles cited in Refs. \cite{PV-z} and \cite{BGMV}, respectively.
One of the authors (K.B.V.) acknowledges useful discussions
with Dr. B. N. Shalaev. This work was supported by the Russian Foundation
for Basic Research via grant No. 01-02-17048, and by the Ministry of
Education of Russian Federation via grant No. E00-3.2-132.

\nid

\end{document}